\documentclass[aps,prd,superscriptaddress,nofootinbib,floats,showpacs]{revtex4-2}
\usepackage{anyfontsize} 

\usepackage{xcolor}
\usepackage{float}
\usepackage{placeins}
\usepackage{stmaryrd}
\usepackage{mathrsfs}
\usepackage{amsmath,amssymb,amsfonts}
\usepackage{graphicx}
\usepackage{subfigure}
\usepackage[normalem]{ulem}
\usepackage{fancyhdr}
\usepackage{hyperref} 
\def\be{\begin{equation}}
	\def\ee{\end{equation}}
\def\ba{\begin{eqnarray}}
	\def\ea{\end{eqnarray}}

\begin{document}

	\title{ Thermodynamic Supercriticality and Complex Phase Diagram for Charged AdS Black Holes in Trace Anomaly Gravity}
	
	\author{Minyan Ou}
	\affiliation{School of Physics and Optoelectronics, South China University of Technology, Guangzhou 510641, China}
	\author{ Xiangdong Zhang\footnote{Corresponding author. scxdzhang@scut.edu.cn}}
	\affiliation{School of Physics and Optoelectronics, South China University of Technology, Guangzhou 510641, China}
	\author{ Zhang-Yu Nie\footnote{Corresponding author. niezy@kust.edu.cn}}
	\affiliation{Center for gravitation and astrophysics, Kunming University of Science and Technology, Kunming 650500, China}


	\begin{abstract}
		We extend the Lee-Yang phase transition framework to charged anti-de Sitter (AdS) black holes in four-dimensional trace anomaly gravity. By treating the horizon radius as a complex variable, we derive a fully resolved complex phase diagram that uncovers novel supercritical phenomena within this modified gravity setting. Relative to the standard Reissner-Nordström-AdS black hole, the trace anomaly shifts the location of the critical point: for a representative set of anomaly parameters, both the critical pressure and critical temperature are suppressed. The Widom line is rigorously identified as the projection of the complex Lee-Yang zeros onto the real physical phase plane, a trajectory that demarcates the small-black-hole-like and large-black-hole-like phases throughout the supercritical regime. We also independently recover this same Widom line via the thermodynamic response function method, demonstrating that the two definitions are in excellent quantitative agreement in the near-critical region and share identical universal scaling behavior. Furthermore, our analysis reveals a smooth, continuous crossover across the Widom line as probed by thermodynamic response functions — a behavior fundamentally distinct from the discontinuous first-order phase transitions that occur below the critical point. These results offer new, physically concrete insights into the thermodynamics of quantum-corrected black holes.
	\end{abstract}
	\maketitle

	\section{Introduction}\label{Intro}
	Since the groundbreaking success of the AdS/CFT correspondence~\cite{maldacena1999large,gubser1998gauge,witten1998anti}-a conjecture identifying the thermodynamics of black holes in anti-de Sitter (AdS) space with that of a dual, strongly coupled conformal field theory (CFT) defined on the AdS boundary-research into AdS black hole thermodynamics has attracted enormous and sustained interest within the high-energy physics community.
	
	Phase transitions represent one of the most fundamental phenomena in thermodynamic systems. Notably, AdS black holes exhibit remarkably rich thermodynamic behaviors, encompassing a diverse range of phase transitions and critical phenomena that have been extensively documented in the literature~\cite{kubizvnak2012p,dayyani2018critical,wei2020extended,lin2024effective,jiang2026higher,hawking1983thermodynamics,Jiang:2026ois}. The paradigmatic example of such behavior is the Hawking–Page phase transition~\cite{hawking1983thermodynamics}, which occurs between a stable large black hole and a thermal gas in AdS space, and is holographically dual to the confinement/deconfinement transition of the quark-gluon plasma on the boundary field theory side \cite{witten1998anti}. These profound correspondences demonstrate that a systematic investigation of black hole thermodynamics in AdS spacetime is not only physically meaningful in its own right, but also provides deep insights into the microscopic nature of black holes and the underlying structure of quantum gravity.
	
	The conformal anomaly occupies a central position in quantum field theory on curved backgrounds. In particular, the trace anomaly is known to play a crucial role in a wide variety of physical processes, including Hawking radiation~\cite{christensen1977trace,robinson2005relationship} and the dynamics of inflationary cosmological models~ \cite{starobinsky1980new,hawking2001trace,nojiri2000brane}. Consequently, exploring the manifestation and consequences of the conformal anomaly in black hole spacetimes has emerged as an important avenue for probing quantum gravitational effects~\cite{cai2010black,cai2014thermodynamics,hu2024quantum}. 
	
	Along this direction, the thermodynamics of AdS black holes with conformal anomaly corrections has previously been systematically analyzed in the literature \cite{cai2014thermodynamics}. It was demonstrated that the inclusion of quantum conformal anomaly terms can qualitatively alter the thermodynamic landscape of these black hole solutions. Most strikingly, the quantum conformal anomaly can induce deviations from the standard scaling laws of ordinary critical phenomena and give rise to entirely new phase structures \cite{hu2024quantum}, which are qualitatively distinct from the conventional van der Waals-like critical behavior widely observed in standard AdS black hole systems.
	
	However, existing investigations into the thermodynamics of AdS black holes modified by the conformal anomaly have predominantly focused on the parameter regime below the critical point, while the possible thermodynamic states above this critical threshold remain comparatively underexplored. For conventional thermodynamic systems ranging from van der Waals fluids to real water, extensive prior studies have firmly established that rich crossover phenomena persist throughout the entire supercritical region~\cite{xu2005relation,simeoni2010widom,ouyang2024complex,wang2025analogous}. This naturally raises a fundamental open question: whether analogous crossover behavior emerges for conformally corrected AdS black holes in their supercritical regime, and if it does, how the quantum trace anomaly reshapes the underlying supercritical thermodynamic structure.
	
	Very recently, Xu et al. extended the applicability of the Lee–Yang phase transition theorem to the framework of AdS black hole thermodynamics, and systematically characterized the supercritical behavior of standard AdS black holes~\cite{xu2026thermodynamic,li2026thermodynamic}. Even though only a single homogeneous thermodynamic phase exists above the critical point, a clear distinction can still be rigorously identified between liquid-like and gas-like dynamical regimes. As the natural continuation of the first-order coexistence curve deep into the supercritical domain, the Widom line acts as a well-defined crossover boundary separating these two distinct regimes. Beyond that, such supercritical crossover phenomena are known to capture the residual signatures of critical fluctuations even in the absence of a genuine phase transition, offering a unique window into the universal thermodynamic properties of strongly interacting systems~\cite{zhao2025characterized,anand2026universal}.
	
	Apart from the Lee–Yang zero formalism, the extrema of thermodynamic response functions provide another widely adopted approach to locate and characterize the Widom line~\cite{xu2005relation,simeoni2010widom,zhao2025characterized,anand2026universal}. Nevertheless, this definition is not strictly unique: the Widom line can be constructed from different thermodynamic response quantities, including the isobaric heat capacity~\cite{xu2005relation,simeoni2010widom} and the isothermal compressibility~\cite{zhao2025characterized}. Recent studies~\cite{zhao2025characterized,anand2026universal} have further demonstrated that dimensionless quantities such as the scaled variance $\Omega$ exhibit sharp, well-resolved extrema across the supercritical region, furnishing an alternative and robust criterion for identifying crossover behavior. Motivated by these recent conceptual advances, in the present work we carry out a detailed investigation of the supercritical thermodynamics of trace-anomaly-corrected AdS black holes from two mutually complementary perspectives, with the central goal of explicitly verifying the consistency between different independent Widom line construction prescriptions.

	The remainder of this paper is organized as follows. In Section 2, we briefly review the basic thermodynamic framework for trace-anomaly-corrected AdS black holes, with a particular focus on their well-established phase behavior below the critical point. Section 3 is dedicated to a systematic investigation of the full thermodynamic structure in the supercritical regime, where we explicitly map out the rich emergent phase landscape that persists beyond the critical threshold. In Section 4, we perform a direct comparative analysis between the Widom lines extracted from two independent construction prescriptions, and quantitatively examine their subtle deviations across the parameter space. Section 5 then focuses on the characteristic crossover behavior in the immediate vicinity of the critical point, clarifying how the trace anomaly modifies the universal scaling properties near the second-order phase transition. Finally, in Section 6, we summarize our main physical results, discuss their possible implications.

	\section{theory}\label{SandW1}
	
	In this section, we provide a concise self-contained review of the trace-anomaly-corrected AdS black hole solution and its key thermodynamic properties. In classical field theories endowed with exact conformal symmetry, the stress-energy tensor is strictly traceless. However, in four-dimensional quantum conformal field theory (CFT), the vacuum expectation value of the trace of the stress-energy tensor no longer vanishes identically \cite{duff1994twenty,deser1993geometric}. This well-established quantum effect is precisely the celebrated trace anomaly, whose explicit form is given by
	\begin{eqnarray}
		\left\langle T^{\mu}_{\mu} \right\rangle_{a} = b I_{4} - a E_{4},
		\label{eq: trace}
	\end{eqnarray}
	
	where \[I_{4} = C_{\mu \nu \lambda \delta} C^{\mu \nu \lambda \delta}, \quad E_{4} = R_{\mu \nu \lambda \delta} R^{\mu \nu \lambda \delta}- 4 R_{\mu \nu} R^{\mu \nu} + R^{2}.\]
	Here, $C_{\mu \nu \lambda \delta}$ denotes the Weyl tensor. The constants $a$ and $b$ are positive coefficients which are related to the degrees of freedom of the underlying quantum field theory \cite{duff1994twenty,deser1993geometric}. The first term in Eq. (\ref{eq: trace}) corresponds to the type B anomaly, while the second term corresponds to the type A anomaly.
	
	Taking the trace anomaly into account, the Einstein equation is modified as follows:
	\begin{eqnarray}
		R_{\mu\nu} - \frac{1}{2} g_{\mu\nu} R + \Lambda g_{\mu\nu} = 8\pi \left\langle T_{\mu\nu} \right\rangle_{a}.
		\label{eq: Einstein equations}
	\end{eqnarray}
	
	When considering a spherically symmetric spacetime with two additional assumptions, namely that the stress energy tensor is covariantly conserved and that
	$\left\langle T^{t}_{t} \right\rangle_{a} = \left\langle T^{r}_{r} \right\rangle_{a}$, one can obtain an exact analytical black hole solution to 
	Eq. (\ref{eq: Einstein equations}) under these assumptions, as shown in \cite{cai2010black,cai2014thermodynamics,hu2024quantum}, which takes the form
	\begin{equation}
		ds^{2} = - f(r)\, dt^{2} + \frac{1}{f(r)}\, dr^{2}
		+ r^{2} \left( d\theta^{2} + \sin^{2}\theta\, d\phi^{2} \right),
	\end{equation}
	where we consider only the type A anomaly and define $\alpha_c = 8\pi a$ and
	\begin{equation}
		f(r) = 1 - \frac{r^{2}}{4 \alpha_c} \left( 1 - \sqrt{ 1 - 8 \alpha_c (\frac{2M}{r^3} - \frac{Q^2}{r^4} - \frac{1}{l^2}) } \right).
		\label{eq: trace bh}
	\end{equation}
	
	Here, we consider only the type A anomaly and define $\alpha_c = 8\pi a$. The $M$ and $Q$ denote the black hole mass and charge, respectively. The cosmological constant is given by \(\Lambda = -\frac{3}{l^{2}}\). In this paper, we focus on the thermodynamic behavior in the supercritical region. Therefore, we restrict our analysis to the AdS case ($\Lambda < 0$).
	
	Using Eq.(\ref{eq: trace bh}), we can express the black hole mass $M$ as a function of the outer horizon radius $r_h$:
	\begin{equation}
		M = \frac{-6 \alpha_c + 3 Q^{2} + 3 r_{h}^{2} + 8 \pi P\, r_{h}^{4}}{6 r_{h}}.
		\label{eq: m}
	\end{equation}
	
	The Hawking temperature is related to the surface gravity at the event horizon, which can be expressed as:
	\begin{eqnarray}
		T = \frac{\mathcal{K}}{2 \pi} = \frac{1}{4 \pi} \frac{\partial f(r)}{\partial r} \Bigg|_{r = r_{h}} = \frac{2 \alpha_c - Q^{2} + r_{h}^{2} + 8 \pi P\, r_{h}^{4}}
		{4 \pi r_{h}^{3} - 16 \alpha_c \pi r_{h}}.
		\label{eq: t}
	\end{eqnarray}
	
	From the Eq.(\ref{eq: t}), we can easily obtain the following relation:
	\begin{eqnarray}
		P = \frac{-2 \alpha_c + Q^{2} - r_{h}^{2} - 16 \alpha_c \pi r_{h} T_{h} + 4 \pi r_{h}^{3} T_{h}}{8 \pi r_{h}^{4}}.
		\label{eq: p}
	\end{eqnarray}
	
	Then the cricitial point of trace-anomaly-corrected AdS black hole satisfied the conditions:
	\begin{eqnarray}
		\left( \frac{\partial P}{\partial V} \right)_{T} = 0,
		\qquad
		\left( \frac{\partial^{2} P}{\partial V^{2}} \right)_{T} = 0.
	\end{eqnarray}
	
	The above equations admit two branches of critical points depending on the value of the trace anomaly parameter $\alpha_c$.
	
\begin{figure}[htbp]
		\centering
		\begin{minipage}[t]{0.45\textwidth}
			\centering
			\includegraphics[width=\textwidth]{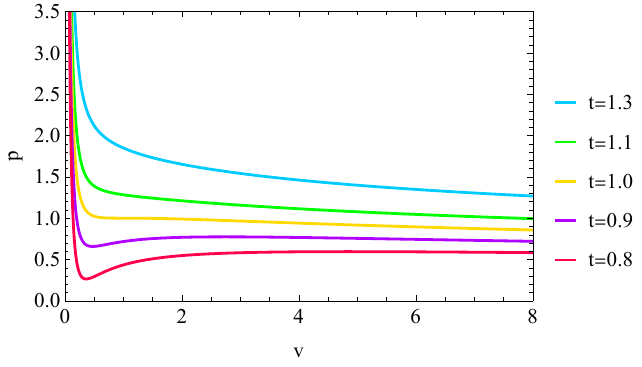}
			\caption{\small   Phase structure of the trace-anomaly-corrected AdS black hole for $\alpha_c <0$. Parameters: $Q=1,\alpha_c=-0.1$.}
			\label{phaseacless8}
		\end{minipage}
		\hfill
		\begin{minipage}[t]{0.45\textwidth}
			\centering
			\includegraphics[width=\textwidth]{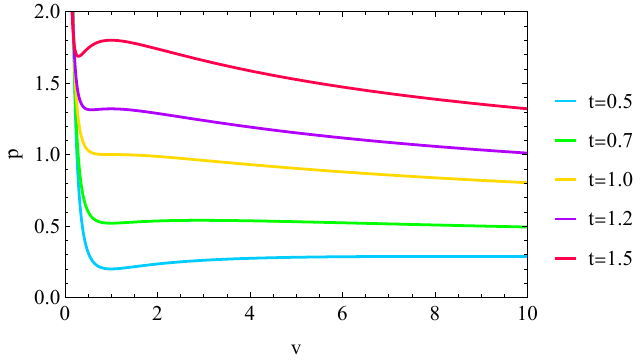}
			\caption{\small Phase structure of the trace-anomaly-corrected AdS black hole for $\alpha_c = \frac{Q^{2}}{8}$. Parameters: $Q=1,\alpha_c=\frac{1}{8}$.}
			\label{phaseac8}
		\end{minipage}
	\end{figure}

	For $\alpha_c \leq 0$, the critical point is given by
	\begin{align}
		V_{c} &= \frac{4\pi}{3} \left(-12 \alpha_c + 3 Q^{2} + K \right)^{3/2}, \\
		T_{c} &= \frac{3 Q^{2} - K}{48 \pi \alpha_c \,\sqrt{-12 \alpha_c + 3 Q^{2} + K} }, \\
		P_{c} &= \frac{-18 \alpha_c + 6 Q^{2} + K}
		{24 \pi \left(-12 \alpha_c + 3 Q^{2} + K \right)^{2}}.
		\label{eq: cricitial point1}
	\end{align}
	where $K \equiv \sqrt{192 \alpha_c^{2} + 9 Q^{4} - 96 \alpha_c Q^{2} }, V_{c} = \frac{4}{3} \pi r_c^3$.
	In the limit $\alpha_c\rightarrow0$, these critical quantities reduce to those of the RN--AdS black hole. The corresponding dimensionless isothermal curves are shown in Fig.~\ref{phaseacless8}. A first-order phase transition occurs only when the temperature is below the critical temperature.
	
	For $0<\alpha_c\leq\frac{Q^{2}}{8}$, the second branch of critical points read
	\begin{align}
		V_{c} &= \frac{4\pi}{3} \left(-12 \alpha_c + 3 Q^{2} - K \right)^{3/2}, \\
		T_{c} &= \frac{3 Q^{2} + K}{48 \pi \alpha_c \, \sqrt{-12 \alpha_c + 3 Q^{2} - K} }, \\
		P_{c} &= \frac{-18 \alpha_c + 6 Q^{2} - K}
		{24 \pi \left(-12 \alpha_c + 3 Q^{2} - K \right)^{2}}.
		\label{eq: cricitial point2}
	\end{align}
	which becomes divergent in the limit $\alpha_c\rightarrow0$.
	
	Remarkably, in the parameter regime $0<\alpha_c<Q^2/8$, the conventional van der Waals-like first-order small/large black-hole phase transition disappears and is replaced by a Hawking--Page phase transition. Although the criticality conditions admit formal solutions, none of them simultaneously satisfies the physical requirements $r_c>0$, $T_c>0$, and $P_c>0$. Consequently, no physically admissible critical point exists in this regime. 
	
	
	The special fine-tuned case $\alpha_c=\frac{Q^{2}}{8}$is explicitly illustrated in Fig.~\ref{phaseac8}. In this particular scenario, first-order phase transitions persist on both sides of the critical temperature, a nonstandard behavior that was first identified and analyzed in Ref.~\cite{hu2024quantum}. This highly anomalous phase structure can be directly traced back to the characteristic violation of the standard mean-field scaling laws induced by the trace anomaly correction. Taken together, these results unambiguously demonstrate that the trace anomaly parameter $\alpha_c$ is not merely a small perturbative correction, but qualitatively reshapes the global thermodynamic phase structure of the AdS black hole system.
	
	
	
	\section{Thermodynamic Supercriticality}\label{SandW2}
	In the previous section, we briefly reviewed the thermodynamic properties and phase structure of the trace-anomaly-corrected AdS black hole near and below the critical point. Similar to the van der Waals fluid system, the RN-AdS black hole exhibits a first-order phase transition between the small black hole and large black hole phases below the critical point \cite{kubizvnak2012p}. The coexistence line separates the small black hole phase from the large black hole phase and terminates at the critical point. Along the coexistence line, thermodynamic quantities such as the Gibbs free energy remain continuous but become non-analytic at the phase transition point. Therefore, the coexistence line plays an important role in characterizing the thermodynamic phase structure of thermodynamic systems.
	
	Above the critical point, the first-order phase transition disappears and the system enters the supercritical region, where have only single thermodynamic phase. Nevertheless, the absence of a true phase transition does not imply the absence of thermodynamic crossover behavior. For conventional thermodynamic systems, such as van der Waals fluids and water systems, extensive studies have shown that rich crossover phenomena still persist in the supercritical region \cite{xu2005relation,simeoni2010widom,ouyang2024complex,wang2025analogous}. 
	
	
	
	Recently, the study of supercritical behavior has been extended to black hole thermodynamics ~\cite{xu2026thermodynamic,li2026thermodynamic,zhao2025characterized,anand2026universal}. Based on the Lee--Yang phase transition theorem, Xu \textit{et al}. demonstrated that the projection of complex Lee--Yang zeros onto the real thermodynamic phase plane is closely related to the Widom line~\cite{xu2026thermodynamic}. As an extension of the coexistence curve into the supercritical region, the Widom line serves as the crossover line separating two thermodynamic regimes and provides a useful framework for characterizing supercritical crossover behavior in black hole systems.
	
	Motivated by the above discussion, we now start to investigate the supercritical thermodynamic behavior of the trace-anomaly-corrected AdS black hole. In this regime, although the system no longer exhibits a phase transition, the thermodynamic crossover behavior remains strongly dependent on the trace anomaly parameter $\alpha_c$. 
       
   	When $\alpha_c=\frac{Q^2}{8}$, as discussed in the previous section and demonstrated in Ref.~\cite{hu2024quantum}, the conventional scaling laws are violated, and first-order phase transitions occur both above and below the critical temperature. Therefore, no supercritical region exists in this special case. 
   	For $0<\alpha_c<Q^2/8$, as established in the previous section, no physically admissible critical point exists. Hence, no associated supercritical region can be defined, and this parameter regime is excluded from the following supercritical analysis.
   
   The following analysis is therefore restricted to $\alpha_c \leq 0$, for which the first-order phase transition terminates at a well-defined critical point and a well-defined supercritical regime exists. Following the approach adopted in Ref.~\cite{xu2026thermodynamic}, we apply the Lee-Yang phase transition theorem to investigate the supercritical thermodynamic behavior of the trace-anomaly-corrected AdS black hole.
       
	Accordingly, the first law of black hole thermodynamics\cite{dolan2011pressure} can be written as
	\begin{eqnarray}
		dU = TdS - PdV + \Phi dQ.
		\label{eq: first law}
	\end{eqnarray}
	where $U$ is the internal energy, $T$ is the Hawking temperature, $S$ denotes the entropy, $P$ is the thermodynamic pressure, $V$ is the thermodynamic volume, and $\Phi$ is the electric potential.
	
	From the first law of thermodynamics, the entropy can be obtained as
	\begin{eqnarray}
		S = \int \frac{dM}{T} = \pi r_h^2 - 4 \pi a_c \log\left(\frac{4 \pi r_h^2}{S_0}\right),
		\label{eq: s}
	\end{eqnarray}
	where $S_0$ is an integration constant.
	
	The minimal horizon radius of the black hole is given by
	\begin{eqnarray}
		r_h = \frac{\sqrt{\frac{-1 + \sqrt{1 - 64 \pi a_c P + 32 \pi P Q^2}}{P \pi}}}{4}.
		\label{eq: mini_horizon}
	\end{eqnarray}
	which ensures a positive Hawking temperature. At this limiting point, the black hole approaches its minimal allowed horizon size, and it is natural to impose that the entropy vanishes in this extremal limit.
	
	Accordingly, one may fix the integration constant $S_0$ as
	\begin{eqnarray}
		S_0 = \frac{e^{-\frac{-\frac{1}{P} + \frac{\sqrt{1 - 64 \pi a_c P + 32 \pi P Q^2}}{P}}{64 \pi a_c}} \left(-1 + \sqrt{1 - 64 \pi a_c P + 32 \pi P Q^2}\right)}{4 P}.
		\label{eq: S0}
	\end{eqnarray}
	

	
	The heat capacity at constant pressure is given by: 
	\begin{eqnarray}
		C_p = T \left( \frac{\partial S}{\partial T} \right)_{P} = - T \left( \frac{\partial^{2} G}{\partial T^{2}} \right)_{P} = -\frac{2 \pi \left(r_h^2 - 4 a_c\right)^2 \left(2 a_c - Q^2 + r_h^2 + 8 \pi P r_h^4\right)}{-8 a_c^2 - 3 Q^2 r_h^2 + r_h^4 - 8 \pi P r_h^6 + 2 a_c \left(2 Q^2 + 5 r_h^2 + 48 \pi P r_h^4\right)}.
		\label{eq: C_p}
	\end{eqnarray}
	where we use the relation $dG = - S\, dT + V\, dP + \Phi\, dQ$.
	
	To simplify the thermodynamic analysis, we introduce the following dimensionless reduced parameters:
	\[p = \frac{P}{P_c}, \quad t = \frac{T}{T_c}, \quad z = \frac{r_h}{r_c}, \quad g = \frac{G}{G_c}, \quad c_p = C_p P_c .\]
	
	Then, the reduced thermodynamic quantities can be written as
	\begin{equation}
		c_p = \frac{\mathcal{N}}{\mathcal{D}}.
		\label{eq: reduced c_p}
	\end{equation}
	where the numerator $\mathcal{N}$ is defined as:
	\begin{equation}
		\begin{aligned}
			\mathcal{N} = {} & \Big( -18 \alpha_c + 6 Q^2 + \Delta \Big) \bigg( \big(3 Q^2 + \Delta\big) z^2 - 4 \alpha_c \big(1 + 3 z^2\big) \bigg)^2 \\
			& \times \Bigg( \Delta z^2 \big(3 + p z^2\big) - 6 \alpha_c \big(-1 + 6 z^2 + 3 p z^4\big) + Q^2 \big(-3 + 9 z^2 + 6 p z^4\big) \Bigg).
		\end{aligned}
	\end{equation}
	and the denominator $\mathcal{D}$ is defined as:
	\begin{equation}
		\begin{aligned}
			\mathcal{D} = {} & 12 \Big( -12 \alpha_c + 3 Q^2 + \Delta \Big)^2 \Bigg( 9 Q^2 \big(3 Q^2 + \Delta\big) z^2 \big(1 - 2 z^2 + p z^4\big) \\
			& + 24 \alpha_c^2 \bigg( 1 + 15 z^2 + \Big(9 p - 42\Big) z^4 + 17 p z^6 \bigg) \\
			& - 6 \alpha_c \bigg( \Delta z^2 \Big(5 + 2\big(p - 6\big) z^2 + 5 p z^4\Big) + Q^2 \Big(2 + 33 z^2 + 12\big(p - 7\big) z^4 + 37 p z^6\Big) \bigg) \Bigg).
		\end{aligned}
	\end{equation}
	with the definition:
	\begin{equation}
		\Delta = \sqrt{192 \alpha_c^2 - 96 \alpha_c Q^2 + 9 Q^4}.
	\end{equation}
	
	
	
	From the Lee-Yang theory of phase transitions, it is known that the behavior of the zeros of the grand partition function governs the occurrence of phase transitions, and these zeros are referred to as Lee-Yang zeros. Fisher further extended this framework to the canonical ensemble \cite{fisher1965nature}.
	
	Within the Euclidean approach to quantum gravity, the Gibbs free energy is related to the partition function through $G = T I = -T \ln Z$, where $G$ denotes the Gibbs free energy, $T$ is the temperature of the canonical ensemble, $I$ is the Euclidean action, and $Z$ is the partition function. Provided that the temperature remains well-defined, the zeros of the partition function correspond to singularities of the Gibbs free energy.
	
	The Gibbs free energy as a function of temperature exhibits a characteristic swallowtail structure, indicating the presence of first-order phase transitions. The two distinct sharp points within the swallowtail structure correspond to non-analytic behavior, where the second derivative of the Gibbs free energy with respect to temperature becomes discontinuous.
	
	From Eq.(\ref{eq: C_p}), we can see that the heat capacity at constant pressure $C_p$ is related to \( \left( \frac{\partial^{2} G}{\partial T^{2}} \right)_{P}.\)
	Therefore, the singular behavior of the heat capacity corresponds to the non-analytic behavior of the Gibbs free energy.
	Consequently, the Lee-Yang zeros of the system can be determined from Eq.(\ref{eq: reduced c_p}), which characterize the singularity structure of the Gibbs free energy:
	\begin{equation}
		\mathcal{D}=0.
		\label{eq: Lee-Yang zeros}
	\end{equation}
	In the following, we treat the variable $p$ to be real, while the variable $z$ is extended into the complex plane, which generally renders $t$ complex \cite{xu2024generalized}. 
	
\begin{figure}[!htb]
		\centering
		\includegraphics[width=0.7\textwidth]{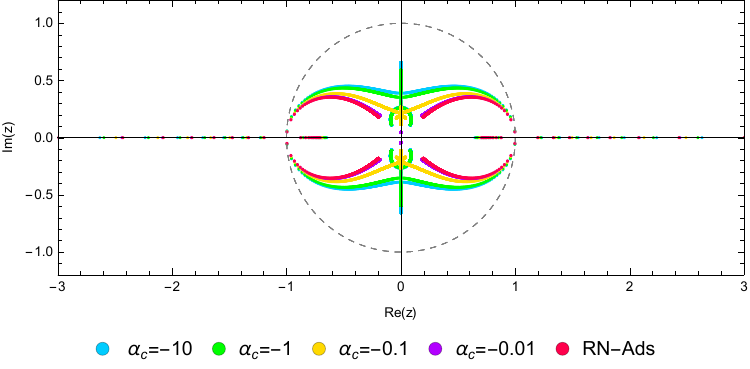}
		\caption{\small Distribution of the Lee--Yang zeros, corresponding to the singularities of the 
			Gibbs free energy, for the trace-anomaly-corrected AdS black hole with different 
			values of the trace anomaly parameter $\alpha_c$ and pressure. Parameters: $Q=1$.}
		\label{naccomplex}
	\end{figure}

\begin{figure}[!htb]
		\centering
		\includegraphics[width=0.7\textwidth]{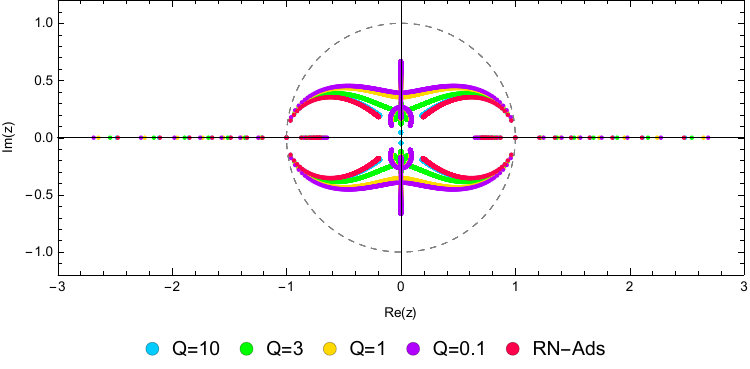}
		\caption{\small Distribution of the Lee--Yang zeros, corresponding to the singularities of the 
			Gibbs free energy, for the trace-anomaly-corrected AdS black hole with different 
			values of the charge $Q$ and pressure. Parameters: $\alpha_c=-1$.}
		\label{nacQcomplex}
	\end{figure}

	In this work, we investigate how the distribution of these singularities depends on the trace anomaly parameter $\alpha_c$ and the black hole charge $Q$. Figures \ref{naccomplex} and \ref{nacQcomplex} display the resulting distributions for different values of $p$.
	
	Since zeros with negative real or imaginary parts of $z$ correspond to unphysical temperatures with negative real or complex values in the thermodynamic interpretation, we restrict our analysis to the first quadrant of the complex plane.
	
	In Fig.\ref{naccomplex}, we fix $Q=1$ and show the effect of the trace anomaly parameter $\alpha_c$ on the distribution of Lee--Yang zeros. For different values of $\alpha_c$, we find that, except for the zeros on the real axis, all other zeros lie inside the unit circle, in agreement with the Lee--Yang unit circle theorem. The zeros on the real axis correspond to genuine thermodynamic phase transitions.
	
	When $p < 1$, the zeros lie on the real axis together with a pair of purely imaginary zeros, which do not correspond to physically meaningful thermodynamic states and are therefore excluded from physical interpretation. When $p > 1$, all nontrivial zeros move off the real axis and extend into the complex plane, while only the fixed point $(1,0)$ remains on the real axis and persists for all parameter values.
	
	As the magnitude of $\alpha_c$ decreases, the complex phase structure gradually approaches that of the RN--AdS black hole. In the limit $\alpha_c \to 0$, both the metric function and the critical point reduce to those of the RN--AdS spacetime. Correspondingly, the distribution of Lee--Yang zeros continuously approaches the RN--AdS configuration, confirming the consistency of the complex phase structure with the RN--AdS limit of the thermodynamic system.
	
	Following the prescription proposed in \cite{xu2026thermodynamic}, the complex phase line determined by the singularity structure of the Gibbs free energy in the first quadrant of the complex plane can be projected onto the real thermodynamic phase plane, yielding the Widom line.
	
\begin{figure}[htbp]
		\centering
		\begin{minipage}[t]{0.45\textwidth}
			\centering
			\includegraphics[width=\textwidth]{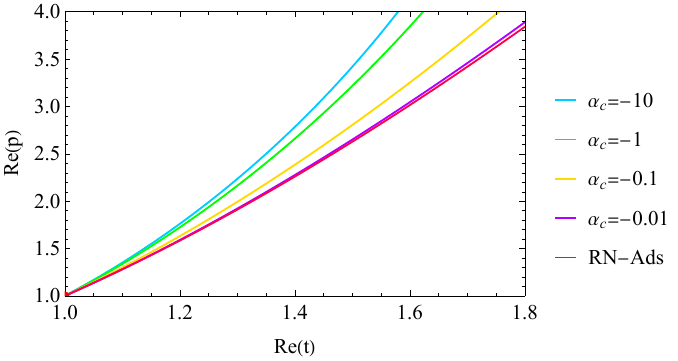}
			\caption{\small The Widom lines determined from the Lee--Yang zeros for the trace-anomaly-corrected AdS black hole with different values of the trace anomaly parameter $\alpha_c$. Parameters: $Q=1$.}
			\label{nacwidom}
		\end{minipage}
		\hfill
		\begin{minipage}[t]{0.45\textwidth}
			\centering
			\includegraphics[width=\textwidth]{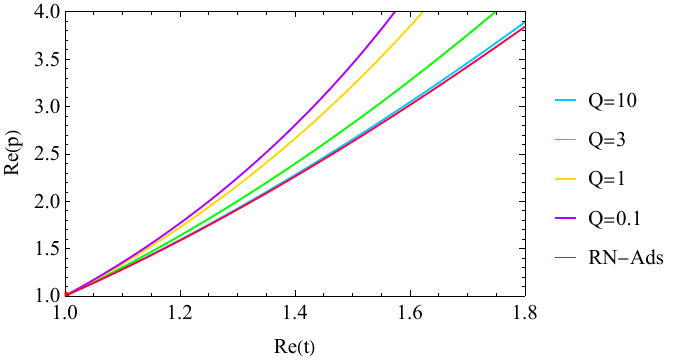}
			\caption{\small The Widom lines determined from the Lee--Yang zeros for the trace-anomaly-corrected AdS black hole with different values of the charge $Q$. Parameters: $\alpha_c=-1$.}
			\label{nacQwidom}
		\end{minipage}
	\end{figure}

	The Widom line extracted from Fig. \ref{naccomplex} is shown in Fig. \ref{nacwidom}. It separates the supercritical region into small black hole-like and large black hole-like regimes. As the magnitude of $\alpha_c$ decreases, the small black hole-like region expands while the large black hole-like region shrinks, and the Widom line gradually approaches that of the RN--AdS black hole.
	
	The corresponding complex phase structure and Widom line for different values of $Q$ are shown in Fig. \ref{nacQcomplex} and Fig. \ref{nacQwidom}, respectively. We find that the effect of the black hole charge $Q$ is qualitatively similar to that of the trace anomaly parameter $\alpha_c$. As $Q$ increases, both the Lee--Yang zero distribution and the Widom line gradually approach the RN--AdS case.
	
	\section{Comparison between Lee-Yang and response-function Widom lines}\label{SandW3}

	Besides the Lee–Yang zero approach, extrema of thermodynamic response functions can also be used to characterize the Widom line \cite{xu2005relation,simeoni2010widom,zhao2025characterized,anand2026universal}. However, its definition is not unique, as it can be constructed from different response functions, such as the heat capacity at constant pressure \cite{xu2005relation,simeoni2010widom} and the isothermal compressibility \cite{zhao2025characterized}. Previous studies\cite{zhao2025characterized,anand2026universal} have shown that quantities such as the scaled variance $\Omega$ exhibit pronounced extrema in the supercritical region, providing an alternative but widely used criterion for defining crossover behavior. Motivated by these developments, we investigate the supercritical behavior of the trace-anomaly-corrected AdS black hole from two complementary perspectives, with the aim of testing the consistency between different Widom line constructions.
	
	Specifically, we compare the Widom line obtained from the projection of Lee–Yang zeros onto the real thermodynamic phase plane with that defined by the extrema of thermodynamic response functions. In this work, we adopt the scaled variance $\Omega$ as a representative response function, defined as:
	\begin{eqnarray}
		\Omega = \frac{k_2}{k_1},
		\label{eq: Omega}
	\end{eqnarray}
	where $k_n \equiv \left( \frac{\partial^n G(T, P)}{\partial T^n} \right)_P$.
	
	In the supercritical region, the Widom line is identified as the locus satisfying $(\partial \Omega / \partial T)_P = 0$ at fixed pressure. 
	
\begin{figure}[htbp]
		\centering
		\begin{minipage}[t]{0.45\textwidth}
			\centering
			\includegraphics[width=\textwidth]{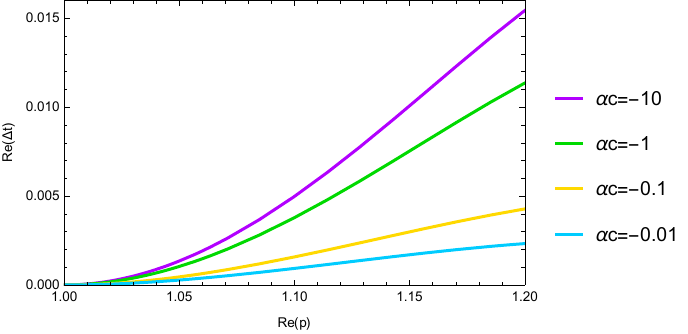}
			\caption{\small Comparison of two methods for determining the Widom line for the 
				trace-anomaly-corrected AdS black hole with different values of the 
				trace anomaly parameter $\alpha_c$. Parameter: $Q=1$.}
			\label{twoacwidom}
		\end{minipage}
		\hfill
		\begin{minipage}[t]{0.45\textwidth}
			\centering
			\includegraphics[width=\textwidth]{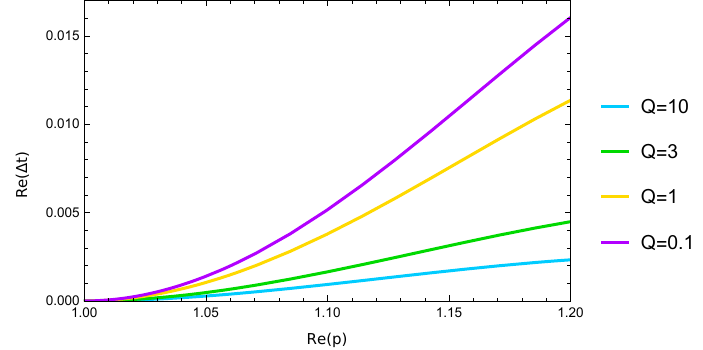}
			\caption{\small Comparison of two methods for determining the Widom line for the 
				trace-anomaly-corrected AdS black hole with different values of the 
				charge $Q$. Parameter: $\alpha_c=-1$.}
			\label{twoQwidom}
		\end{minipage}
	\end{figure}

	To quantify the agreement between the two constructions, we define the temperature difference $\Delta t = t_{\mathrm{LY}} - t_{\Omega}$ at the same pressure. As illustrated in Fig. \ref{twoacwidom} and Fig. \ref{twoQwidom}, the two Widom line definitions coincide in the vicinity of the critical point, where $\Delta t$ is strongly suppressed and approaches zero across all parameter choices. This agreement indicates that, near criticality, both constructions are governed by the same universal scaling behavior, and therefore yield an equivalent characterization of the crossover structure.
	
	Away from the critical point, a gradual deviation between the two Widom line definitions becomes observable. This behavior reflects the increasing sensitivity of the crossover structure to the choice of thermodynamic probe in the deep supercritical regime, where universality progressively weakens and observable-dependent features begin to emerge. In addition, the magnitude of this deviation is affected by system parameters: increasing the trace anomaly parameter $\alpha_c$ or decreasing the black hole charge $Q$ tends to enhance the discrepancy between the two constructions.
	
	Overall, Fig. \ref{twoacwidom} and Fig. \ref{twoQwidom} demonstrates that the two Widom line constructions are fully consistent in the critical region, while mild deviations appear only away from criticality.
	
	\section{Supercritical crossover}\label{SandW4}
	
	Trace-anomaly-corrected AdS black hole exhibit a first-order small/large black hole phase transition below the critical point when $\alpha_c \leq 0$. Above the critical point, the system undergoes a smooth and continuous supercritical crossover. This set of supercritical crossover lines $L_\pm$ can serve as sharp probes for 
	critical fluctuations and mean-field scaling laws. They not only quantitatively reveal the extended behavior of critical dynamics within the supercritical region, but also provide a clear and rigorous pathway for 
	characterizing the fluid-like behavior of black hole thermodynamics under the supercritical regime.
	
	In order to get the crossover lines $L_\pm$, we first need to get the Widom line near the critical point. Here we use the extrema of the scaled variance $\Omega$ to dertemine the Widom line, as the set $\alpha_c=-0.1, Q=1$, we can nummerical solve the $(\partial \Omega / \partial T)_P = 0$ and ger the Widom line. Introduce the dimensionless reduced thermodynamic variables $\Delta T = \frac{T}{T_c} - 1$ and $\Delta P = \frac{P}{P_c} - 1$, and expanding around the critical point using the critical quantities, we can get the:
	\begin{equation}
		\Delta P = 2.832 \Delta T + 1.457 \Delta T^2 + 0.173 \Delta T^3 + \dots.
		\label{eq:widom near}
	\end{equation}
	
	To obtain the crossover lines $L_\pm$ in the supercritical region, we introduce an order parameter within the extended ensemble. Typically, the density variable $\rho$ is selected as the order parameter, where $\rho = 1/v$ and 
	$v$ denotes the specific volume. 
	
	At the critical point,
	\begin{equation}
		\left. \frac{\partial P}{\partial \rho} \right|_c = 0, \quad \left. \frac{\partial^2 P}{\partial \rho^2} \right|_c = 0.
		\label{eq:state}
	\end{equation}
	
	Introduce the order parameter deviation $m = \rho - \rho_c$ to quantify the degree of fluctuations deviating from the critical density. Within this framework, the pressure deviation $\delta P$, acting as the conjugate field of the order parameter, can be described near the critical point via a Landau expansion of the equation of state Eq. (\ref{eq: p}) then give us:
	\begin{equation}
		\delta P \equiv P(\rho, T) - P(\rho_c, T) = A m \tau + B m^3 + \dots.
	\end{equation}
	where the temperature deviation is defined as $\tau = T - T_c$, and the expansion coefficients $A$ and $B$ are determined by the partial derivatives at the critical point, respectively:
	\begin{equation}
		A = \left. \frac{\partial^2 P}{\partial \rho \partial T} \right|_c , \quad B = \left. \frac{1}{6} \frac{\partial^3 P}{\partial \rho^3} \right|_c.
	\end{equation}
	
	The density fluctuation behavior of the system can be quantitatively characterized by the isothermal susceptibility $\chi_T$. Based on the Landau expansion of the equation of state near the critical point, this susceptibility 
	is physically defined as:
	\begin{equation}
		\chi_T = \left( \frac{\partial m}{\partial \delta P} \right)_T = \frac{1}{A \tau + 3 B m^2}.
		\label{eq:xt}
	\end{equation}
	This physical quantity quantifies the sensitivity of the system's order parameter to small perturbations of external pressure at a constant temperature. To locate the boundaries with the most intense fluctuations within the supercritical region, we impose the extremum condition on the isothermal susceptibility, namely $\partial_m \chi_T = 0$. Furthermore, utilizing Eq. (\ref{eq:xt}), finally yields the two supercritical crossover lines for the trace-anomaly-corrected AdS black hole:
	\begin{equation}
		m_\pm = \pm \sqrt{\frac{A}{3B} \tau}, \quad \delta P_\pm = \pm \frac{4}{3\sqrt{3}} \frac{A^{3/2}}{B^{1/2}} \tau^{3/2}.
	\end{equation}
	In the physical picture, these two bifurcated branches, $m_+$ and $m_-$, correspond to the system's fluctuations toward the high-density or low-density regions relative to the critical density $\rho_c$, respectively. 
	
	Furthermore, we introduce the dimensionless reduced thermodynamic variables $\Delta T = \frac{T}{T_c} - 1$ and $\Delta P = \frac{P}{P_c} - 1$. After defining the reduced pressure deviation $\Delta P_{\pm} = \frac{\delta P_{\pm}}{P_c}$, the aforementioned two supercritical crossover lines obey the following relationship:
	\begin{equation}
		\Delta P_{\pm}(\Delta T) =3.661 \Delta T^{3/2} + \dots.
		\label{eq: fluctuations}
	\end{equation}
	
	Combining Eq. (\ref{eq:widom near}) with Eq. (\ref{eq: fluctuations}), we obtain the universal form of the crossover lines:
	\begin{equation}
		\Delta P_{L\pm}(\Delta T) =  2.832 \Delta T \pm 3.661 \Delta T^{3/2} + \dots.
	\end{equation}
	
\begin{figure}[!htb]
		\centering
		\includegraphics[width=0.6\textwidth]{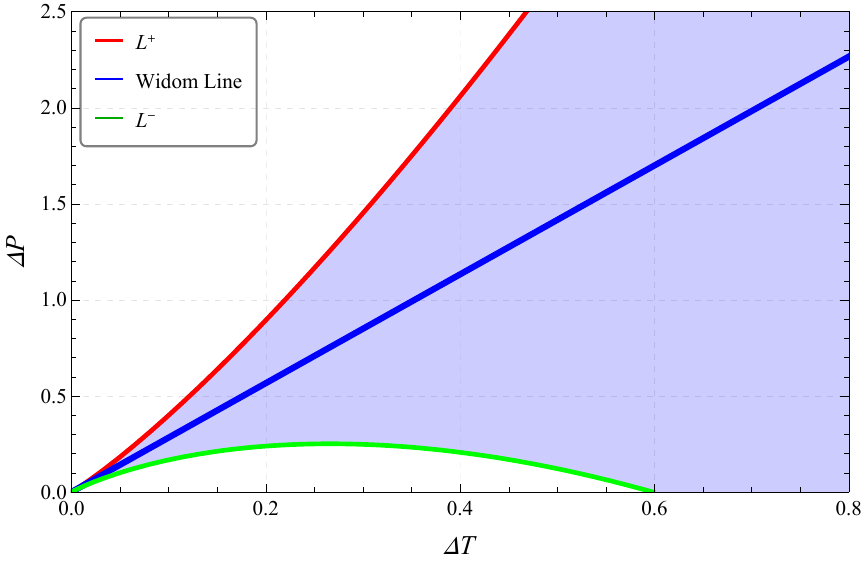}
		\caption{\small Supercritical crossover structure in the $P$--$T$ plane for the 
			trace-anomaly-corrected AdS black hole. Parameters: $\alpha_c=-0.1, Q=1$.}
		\label{cross}
	\end{figure}

	As the Fig. (\ref{cross}) show that, the corresponding $\Delta P_{L_{\pm}}(\Delta T)$ curves illustrate the thermodynamic structure of the black hole in the supercritical regime. The central Widom line $P(T)$ reflects the mean-field trajectory in the vicinity of the critical point, corresponding to the locus of maximal correlation length and providing a natural continuous extension of the first-order coexistence line into the supercritical region.
	
	The nonlinear branches $\Delta P_{L\pm}(\Delta T)$ characterize deviations from this central behavior and define an envelope associated with enhanced thermodynamic response. The geometric structures of both the central trajectory and its bounding curves are explicitly influenced by the trace anomaly parameter $\alpha_c$ and the black hole charge $Q$. The region enclosed by $P_{L+}(T)$ and $P_{L-}(T)$ forms a crossover fan, within which various thermodynamic response functions exhibit pronounced extrema.
	
	\section{ CONCLUSIONS AND DISCUSSIONS}\label{SandW5}
	
	This work presents a systematic investigation of the full thermodynamic structure of trace-anomaly-corrected AdS black holes deep within the supercritical regime. Building upon a complete characterization of the phase diagram below the critical point, we analytically extend the horizon radius to a complex variable, and systematically explore the singularity structure of the Gibbs free energy in the extended complex phase space within the rigorous framework of Lee–Yang phase transition theory. This approach allows us to explicitly map out the universal analytic features that govern the entire supercritical region.
	
	Since the trace anomaly coefficient $\alpha_c$ acts as a non-trivial control parameter that qualitatively modulates the thermodynamic behavior of the system, we focus on the physically well-defined supercritical regime with $\alpha_c \le 0$. In this regime, as the magnitude of $\alpha_c$ decreases or the black hole charge $Q$ increases, the distribution of Lee–Yang zeros and the associated Widom line smoothly converge toward the well-established limiting behavior of the standard Reissner–Nordström-AdS (RN-AdS) black hole.
	
	Furthermore, we perform a direct quantitative comparison between two independent prescriptions for constructing the Widom line: the first is defined via the projection of Lee–Yang zeros onto the physical thermodynamic plane, while the second is extracted from the extrema of the scaled variance $\Omega$ within the conventional thermodynamic response-function formalism. Our numerical results demonstrate that these two distinct definitions exhibit excellent quantitative agreement in the immediate neighborhood of the critical point, faithfully capturing the same universal scaling behavior. However, they develop a gradual, measurable deviation as one moves far away from the critical point deep into the supercritical regime. This observed deviation explicitly demonstrates that the Widom line is not a uniquely defined object in black hole thermodynamics, and highlights the non-trivial role of the quantum trace anomaly in reshaping the fine structure of supercritical crossover phenomena.
	
	Finally, by performing a systematic Landau-type expansion of the equation of state around the critical point, we derive the analytic form of the crossover lines $\Delta P_{L\pm}(\Delta T)$ in the supercritical region. While their functional form remains universal across different parameter values, the corresponding coefficients explicitly depend on the trace anomaly parameter $\alpha_c$ and the black hole charge Q. The crossover fan region enclosed by these two lines precisely marks the parameter domain where thermodynamic response functions are strongly enhanced, thus providing a clear geometric characterization of the regime dominated by intense supercritical fluctuations.
	
	Taken together, this work establishes a unified physical picture in which the trace anomaly parameter $\alpha_c$ and the black hole charge $Q$ act as two independent control parameters that jointly govern the deformation of the supercritical thermodynamic structure. This deformation is consistently encoded across all complementary observables, including the Lee–Yang zero spectrum, the non-uniquely defined Widom line, and the characteristic geometry of the crossover fan region.
	
	\begin{acknowledgments}
		This work is supported by National Natural Science Foundation of China (NSFC) with Grant No. 12275087 and 12575054. 
	\end{acknowledgments}

	\bibliographystyle{unsrt}
	\bibliography{Ref}

\end{document}